\documentclass[onecolumn,showpacs,preprintnumbers,amsmath,amssymb]{revtex4}

\usepackage{graphicx}
\usepackage{dcolumn}
\usepackage{bm}
\usepackage{epsfig}
\usepackage{graphics}
\usepackage{graphicx}
\usepackage{mathtext}

\begin{document}

\title{
Scale-free coordination number disorder and multifractal size disorder in weighted planar stochastic lattice
}

\author{M. K. Hassan$^{1}$, M. Z. Hassan$^{2}$ and N. I. Pavel$^{1}$}

\affiliation{
$1$  Theoretical Physics Group, Department of Physics, University of Dhaka, Dhaka 1000, Bangladesh \\
$2$ Institute of Computer Science, Bangladesh Atomic Energy Commission, Dhaka 1000, Bangladesh
}

\date{\today}%


\begin{abstract}%

The square lattice is perhaps the simplest cellular structure. In this work, however, we
investigate the various structural and topological properties of the kinetic and stochastic counterpart of the
square lattice and termed them as kinetic square lattice (KSL) and weighted planar stochastic lattice (WPSL) respectively. 
We find that WPSL evolves following several non-trivial conservation laws, 
$\sum_i^N x_i^{n-1} y_i^{{{4}\over{n}}-1}={\rm const.}\  \forall \ n$, where $x_i$ and $y_i$ are the length 
and width of the $i$th block. The KSL, on the other hand, evolves following only one conservation law, 
namely the total area, although one find three apparently
different conserved integrals which effectively the total area.
We show that one of the conserved quantity of the WPSL obtained either by setting $n=1$ or $n=4$ can be used to perform 
multifractal analysis. For instance, we show that if the $i$th block is populated with either $p_i\sim x_i^3$ or 
$p_i\sim y_i^3$ then the resulting distribution in the WPSL exhibits multifractality.
Furthermore, we show that
the dual of the WPSL, obtained by replacing each block with a node at its center and common border between blocks with an edge joining 
the two vertices, emerges as a scale-free network since its degree distribution exhibits power-law $P(k)\sim k^{-\gamma}$ with exponent $\gamma=5.66$.
It implies that the coordination number distribution of the WPSL is scale-free in character 
as we find that $P(k)$ also describes the fraction of blocks having $k$ neighbours.

\end{abstract}


\maketitle



\section{Introduction}

The formation of planar cellular structures have attracted considerable interest among scientists
in general and physicists in particular because cellular structures are ubiquitous in nature. 
Especially, the class of non-equilibrium systems with apparent disordered patterns have come under close scrutiny.
Examples of cellular structures include acicular texture in martensite growth, tessellated pavement on ocean shores,
agricultural land division according to ownership, 
grain texture in polycrystals, cell texture in biology, 
soap froths and so on \cite{ref.martensite, ref.polycrystal,ref.biocell,ref.soapfroths}.
Most existing theoretical models which can mimic such systems are typically formed by tessellation, tiling, or subdivision 
of a plane into domains of contiguous and nonoverlapping cells. 
For instance, Voronoi lattice is formed by partitioning a plane 
into mutually exclusive convex polygons,
 Apollonian packing on the other hand is
formed by tiling a plane into contiguous and non-overlapping disks, etc. These models have 
found widespread applications in physics and biology
\cite{ref.model,ref.apollonian}. Significant progress has already been made in understanding various properties of both
Voronoi lattice and Apollonian packing.

A fact worth mentioning is that most of the existing models for generating cellular structures
have one thing in common and that is: the number of neighbours of a cell never
exceeds the number of sides it has. Another important property of the existing cellular structures is that they have small
mean coordination number where it is almost impossible to find cells that have
significantly higher or fewer neighbours than the average number of neighbours. However, geologists and soil 
scientists observe an 
abundance of cellular patterns ranging from topographical ground formations and underwater stone arrangements
where number of neighbours may exceeds the number of sides of the cell. 
Typically, these structures emerge through evolution and in the process each cell
may acquire more neighbours than the number of sides the corresponding cell has. 
Such disordered structures can be of great interest in physics 
provided they have some global topological and geometrical properties since they can mimic kinetic disordered medium
on which one can study percolation and random walk problems.

\begin{figure}[ht]
\begin{center}
\includegraphics[width=10.5cm,height=9.75cm]{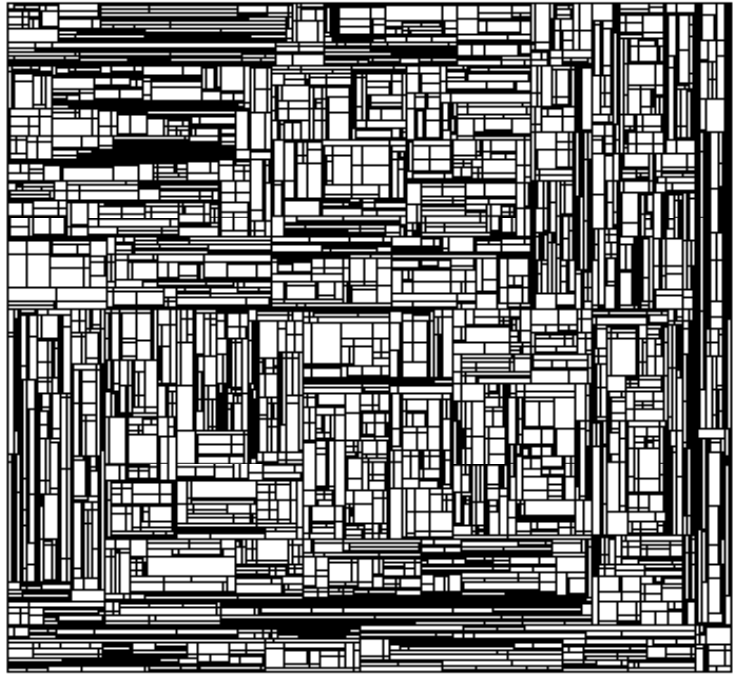}
\end{center}
\caption{A snapshot of the weighted planar stochastic lattice containing $30001$ blocks.
}
\label{fig:fg1}
\end{figure}


Perhaps, the square lattice is the simplest example of the cellular structure where every cell has the same size and 
the same coordination number. Its construction starts with an initiator, say a square of unit area, and a generator 
that divides it into four equal
parts. In the next step and steps thereafter the generator is applied to all the available blocks which eventually 
generates
a square lattice. In this talk, we intend to address the following questions. Firstly, what if the generator is applied
to only one of the available block at each step by picking it preferentially with respect to the areas? Secondly, 
what if we use a modified generator that divides the initiator randomly into four blocks instead of four equal parts and 
apply it to only one of the available block at each step, which are again picked preferentially with respect to their
respective areas \cite{ref.hassan_njp}?  Our primary focus, however, will be on the later case which results in the 
tiling of the initiator into 
increasingly smaller mutually exclusive rectangular blocks. The process is so simple that even a few steps by hand 
on a piece of paper can lead to the conclusion
that the blocks in the resulting lattice may have more neighbours than the number of sides of a cell. We term the resulting 
structure as {\it weighted planar stochastic lattice} (WPSL) since the spatial randomness is incorporated by the modified generator and 
also the time is incorporated in it by the sequential application of the modified generator. 
The definition of the model may appear too simple but the 
results it offers are found to be far from simple. To illustrate the type of
systems expected, we show a snapshot of the resulting weighted planar stochastic lattice taken during the evolution (see figure 1). 
We intend to investigate its topological and geometrical properties 
in an attempt to find some order in this seemingly disordered lattice.

The model in question can also describe the kinetics of fragmentation in two dimensions as it can be defined as 
follows \cite{ref.hassan,ref.krapivsky}. 
At each time step one may assume that a seed is being 
nucleated randomly on the initiator and hence the greater the area of the block, the higher
is the probability that it contains the seed. Upon nucleation two orthogonal cracks parallel to the sides of the block
are grown until intercepted by existing cracks which divides it randomly into four rectangular fragments.  
In reality though, fragments are produced in fracture of solid objects by
propagation of interacting cracks resulting in the fragments of arbitrary size and shape which
makes it a formidable mathematical problem.
However, the present model can be considered as the minimum model which should be capable of capturing the essential 
generic aspects of the underlying mechanism. In addition, Ben-Naim and Krapivsky have noted that the WPSL 
can also describe kinetics of martensite formation. A clear testament to this can be found if one compares figure 1 of our work 
and figure 2 of the  work of Rao {\it et al.} in the context of martensite formation \cite{ref.martensite,ref.martensite_1}. 
Yet another application, albeit a little exotic, is the 
random search tree problem in computer science \cite{ref.majumdar}.

A truly disordered lattice where the coordination number disorder and the block size disorder are introduced naturally
is perhaps long over due. On the other hand, where there is a disorder physicists have the natural tendency to 
look for an order in it. To this end, we invoke the idea of multifractality to quantify the size disorder and the 
idea of scale-invariance to quantify the coordination number disorder. Firstly, we characterize the
blocks of the lattice by the size of their respective length $x$ and width $y$. Furthermore,
we identify each blocks by labelling them as $i=1,2,...,N$ and the corresponding length 
and width by
$(x_1,y_1)$, $(x_2,y_2)$, $...$, $(x_N,y_N)$ respectively. We then show that the dynamics of the WPSL is governed by infinitely many conservation laws, namely the numerical 
value of $\sum_i^N x_i^{n-1}y_i^{4/n-1}$ remains the same regardless of the lattice size $N$ and the value of $n$
where $n=2$ corresponds to the conservation of total area. 
Of the non-trivial conservation laws, we single out either $\sum_i^N y_i^3$ by setting $n=1$ or 
$\sum_i^N x_i^3$ by setting $n=4$ as we can use it to perform multifractal analysis. For instance, we show that if the blocks are populated according to 
the cubic power of their own length (or width) then the distribution of the population in the WPSL 
emerges as multifractal. On the other hand, if the blocks are characterized by the coordination number or the number of neighbours $k$ with whom
the block have common sides then we show that the coordination number distribution function $\rho(k)$, the probability
that a block picked at random have coordination number $k$, have power-law tail.
A power-law distribution function is regarded as scale-free since
it looks the same regardless of the scale we use to look at it.

The organization of this talk is as follows. In section 2, we give the exact algorithm of the model. 
In section 3, we give the geometric properties of the WPSL in an attempt to quantify the annealed size disorder and
at the same time we proposed kinetic square lattice in an attempt to look for a possible origin of multifractality.
In section 4, we discuss the various structural topological 
properties of the WPSL and its dual are discussed in order to quantify the annealed coordination number disorder. 
Finally, section 5 gives a short summary of our results.

\begin{figure}[ht]
\begin{center}
\includegraphics[width=13.5cm,height=6.5cm]{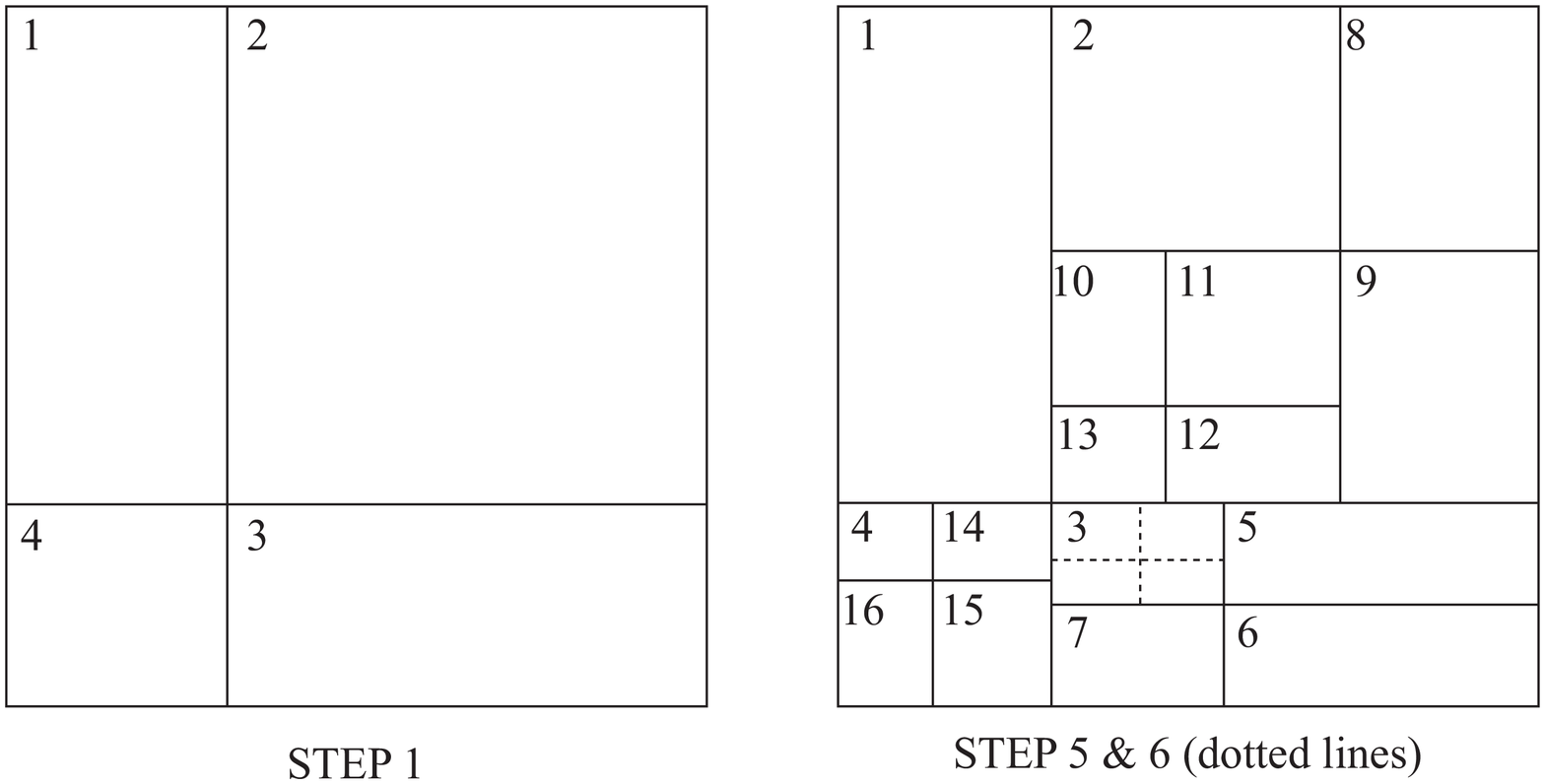}
\end{center}
\caption{Schematic illustration of the first few steps of the algorithm.
}
\label{fig:fg2}
\end{figure}

\section{Algorithm of the weighted planar stochastic lattice (WPSL)}

Perhaps an exact algorithm can provide a better description of the model than the mere definition.
In step one, the generator divides the initiator,
say a square of unit area, randomly into four smaller blocks. The four newly created blocks are then labelled
by their respective areas $a_1, a_2, a_3$ and $a_4$ in a clockwise fashion starting from the upper left block 
(see Fig. 2). In step two and thereafter
only one block is picked at each step with probability equal to their respective area and then it is divided
randomly into four blocks. In general, the $j$th step of the algorithm can be described as follows.
\begin{itemize}

\item[{\bf(i)}]  Subdivide the interval $[0,1]$ into $(3j-2)$ subintervals of size $[0,a_1]$, $[a_1, a_1+a_2]$,$\  ...$, 
$[\sum_{i=1}^{3j-3} a_i,1]$ each of which represent the blocks labelled by their areas $a_1,a_2,...,a_{(3j-2)}$ respectively.

\item[{\bf(ii)}]  Generate a random number $R$ from the interval $[0,1]$ and find which of the $(3i-2)$ sub-interval 
contains this $R$. The corresponding block it represents, say the $p$th block of area $a_p$, is picked.  

\item[{\bf(iii)}] Calculate the length $x_p$ and the width $y_p$ of this block and keep note of the coordinate of the
lower-left corner of the $p$th block, say it is $(x_{low}, y_{low})$.

\item[{\bf(iv)}] Generate two random numbers $x_R$ and $y_R$ from $[0,x_p]$ and $[0,y_p]$ respectively and hence
the point $(x_{R}+x_{low},y_{R}+y_{low})$ mimicking a random nucleation of a seed in the block $p$.

\item[{\bf(v)}] Draw two perpendicular lines through the point $(x_{R}+x_{low},y_{R}+y_{low})$ 
parallel to the sides of the $p$th block mimicking orthogonal cracks parallel to the sides of the blocks
which stops growing upon touching existing cracks and divide it into four smaller blocks. 
The label $a_p$ is now redundant and hence
it can be reused.

\item[{\bf(vi)}]  Label the four newly created blocks according to their areas $a_p$, $a_{(3j-1)}$, $a_{3j}$ and $a_{(3j+1)}$ respectively 
in a clockwise fashion starting from the upper left corner.

\item[{\bf(vii)}] Increase time by one unit and repeat the steps (i) - (vii) {\it ad infinitum}.

\end{itemize}

\section{Geometric properties of WPSL}

In general, the distribution function $C(x,y;t)$ describing the blocks of the lattice by their length $x$ and width $y$ 
evolves according to the following kinetic equation
\cite{ref.hassan,ref.krapivsky}
\begin{eqnarray}
\label{eq:1}
{{\partial C(x,y;t)}\over{\partial t}}& =& -C(x,y;t)\int_0^x\int_0^y dx_1dy_1F(x_1,x-x_1,y_1,y-y_1)
+ \\ \nonumber & & 
4\int_x^\infty\int_y^\infty C(x_1,y_1;t)F(x,x_1-x,y,y_1-y)dx_1dy_1,
\end{eqnarray}
where kernel $F(x_1,x_2,y_1,y_2)$ determines the rules and the rate at which the block of sides $(x_1+x_2)$ and $(y_1+y_2)$
is divided into four smaller blocks whose sides are the arguments of the kernel. The first term on the right hand side of
equation (\ref{eq:1}) represents the loss of blocks of sides $x$ and $y$ due to nucleation of seed of crack on one such block from which
mutually perpendicular cracks are grown to divide it into four smaller blocks. Similarly, the second term on the right 
hand side represents the gain of blocks of sides $x$ 
and $y$ due to nucleation of seed of crack on a block of sides $x_1$ and $y_1$ ensuring 
that one of the four new blocks have sides $x$ and $y$. 
Let us now consider the case where the generator divides the initiator randomly into four smaller rectangles
and apply it to 
only one of the available squares thereafter by picking preferentially with respect to their areas.
It effectively describes the random sequential 
nucleation of seeds with uniform probability on the initiator. Within the rate equation approach this
can be ensured if one chooses the following kernel
\begin{equation}
\label{eq:2}
F(x_1,x_2,y_1,y_2)=1
\end{equation}
Substituting it into equation (\ref{eq:1}) we obtain 
\begin{equation}
\label{eq:3}
{{\partial C(x,y;t)}\over{\partial t}} = -xyC(x,y;t)+  
4\int_x^\infty\int_y^\infty C(x_1,y_1;t)dx_1dy_1.
\end{equation}
The coefficient $xy$ of $C(x,y;t)$ of the loss term implies that seeds of cracks are nucleated on the blocks 
preferentially with respect to their areas which is consistent with the definition of our model.

Incorporating the $2$-tuple Mellin transform given by 
\begin{equation}
\label{eq4}
M(m,n;t)=\int_0^\infty\int_0^\infty x^{m-1}y^{n-1}C(x,y;t)dxdy,
\end{equation}
in equation (\ref{eq:3}) we get
\begin{equation}
\label{eq:momenteq}
{{dM(m,n;t)}\over{dt}}=\Big ( {{4}\over{mn}}-1\Big )M(m+1,n+1;t).
\end{equation}
Iterating equation (\ref{eq:momenteq}) to get all the derivatives of $M(m,n;t)$ and
then substituting them into the Taylor series expansion of $M(m,n;t)$ about $t=0$ one
can immediately write its solution in terms of generalized hypergeometric function \cite{ref.hypergeometric}
\begin{equation}
\label{eq:5}
M(m,n;t)=  ~_2F_2\Big (a_+,a_-;m,n;-t\Big ),
\end{equation}
where $M(m,n;t)=M(n,m;t)$ for symmetry reason and 
\begin{equation}
\label{eq:6}
a_{\pm} = {{m+n}\over{2}} \pm \Big [ \Big ({{m-n}\over{2}}\Big )^2+4 \Big ]^{{{1}\over{2}}}.
\end{equation}  
One can see that (i) $M(1,1;t)=1+3t$ is the total
number of blocks $N(t)$ and (ii) $M(2,2;t)=1$ is the sum of areas of all the blocks
which is obviously a conserved quantity. Both properties are again consistent with the definition of
the WPSL depicted in the algorithm. The behaviour of $M(m,n;t)$ in the long time limit is
\begin{equation}
\label{eq:aminus}
M(m,n;t)\sim t^{-a_-}.
\end{equation}
Thus, in addition to the conservation of total area the system is also governed by infinitely many
non-trivial conservation laws as it implies 
\begin{equation}
\label{eq:7}
M(n,4/n;t)\sim {\rm constant} \hspace{0.25cm}  \ \forall n.
\end{equation}
We used numerical simulation to verify equation (\ref{eq:7}) or its discrete counterpart $\sum_i^N x_i^{n-1}y_i^{4/n-1}$ 
if we label all the available blocks as $i=1,2,...,N$. We found 
that the analytical solution is in perfect agreement with the numerical simulation which we performed based on the 
algorithm for the WPSL model (see figure 3). 

\begin{figure}[ht]
\begin{center}
\includegraphics[width=11.5cm,height=8.5cm]{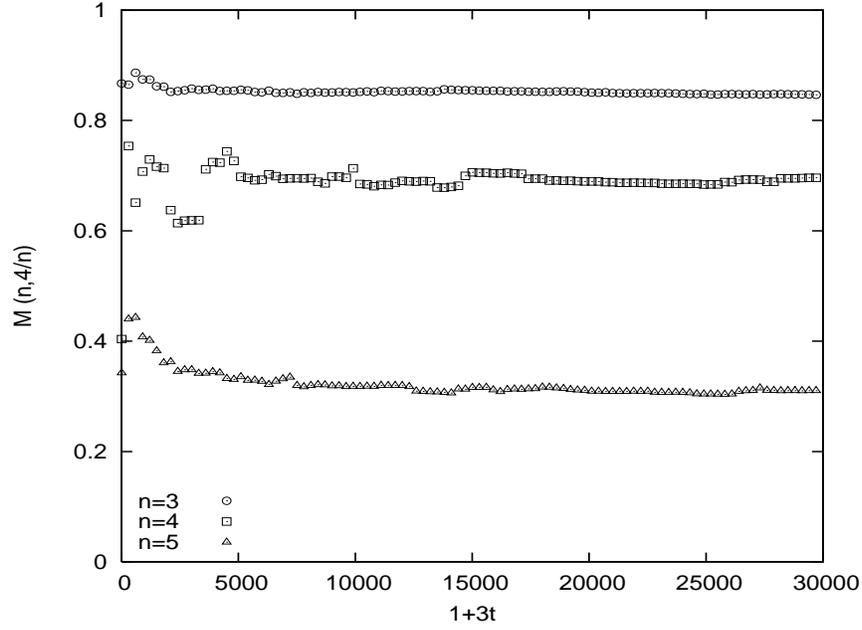}
\end{center}
\caption{The plots of $\sum_i^N x_i^{n-1}y_i^{4/n-1}$ vs $N$ for $n=3,4,5$ are drawn using data collected from
one realization.
}
\label{fig:fg3}
\end{figure}
\begin{figure}[ht]
\begin{center}
\includegraphics[width=12.5cm,height=8.5cm]{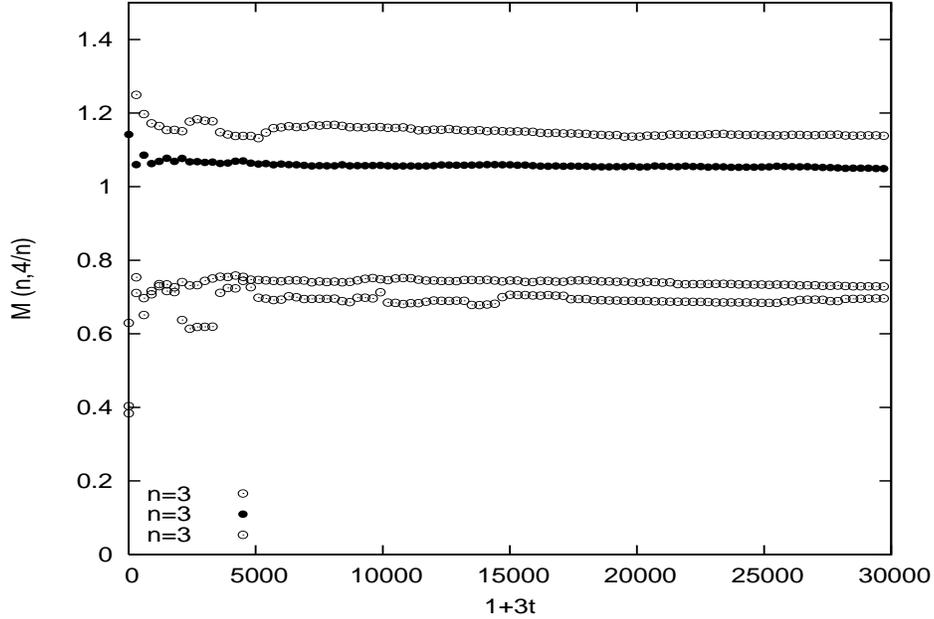}
\end{center}
\caption{The plots of $\sum_i^N x_i^3$ vs $N$ for four different realizations shows that
the numerical value is different at every independent realization.
}
\label{fig:fg4}
\end{figure}

We now find it interesting to focus on the distribution function $n(x,t)=\int_0^\infty C(x,y,t)dy$ 
that describes the concentration of blocks which have length $x$ at time $t$ regardless of the size of their widths $y$. 
Then the $q$th moment of $n(x,t)$ is defined as 
\begin{equation}
\label{eq:8}
M_q(t)=\int_0^\infty x^q n(x,t)dx.
\end{equation}
Appreciating the fact that $M_q(t)=M(q+1,1;t)$ and using equation (\ref{eq:aminus}) one can immediately write its solution
\begin{equation}
\label{eq:qmoment}
M_q(t)\sim t^{\{\sqrt{q^2+16}-(q+2)\}/2}.
\end{equation}
Note that for symmetry reason it does not matter whether we consider the $q$th moments of $n(x,t)$ or that of
$n(y,t)$ since we have $M(q+1,1;t)=M(1,q+1;t)$. 
According to equation (\ref{eq:qmoment}) the quantity $M_3(t)$ and hence $\sum_i^N x_i^3$ or $\sum_i^N y_i^3$ 
is a conserved quantity. 
However, although $M_3(t)$ remains a constant against time in every independent realization 
the exact numerical value is found to be different in every different realization (see figure 4). 
It clearly indicates lack of self-averaging or wild fluctuation. This is supported by our analytical
solution as we find that 
\begin{equation}
\label{eq:9}
<x^q>={{\int_0^\infty x^qn(x,t)dx}\over{\int_0^\infty n(x,t)dx}}\neq <x>^q=
\Big ({{\int_0^\infty xn(x,t)dx}\over{\int_0^\infty n(x,t)dx}}\Big )^q,
\end{equation}
which suggest that a single length scale cannot characterize all the moments of the distribution function $n(x,t)$.

\begin{figure}[ht]
\begin{center}
\includegraphics[width=8.5cm,height=10.5cm,angle=-90]{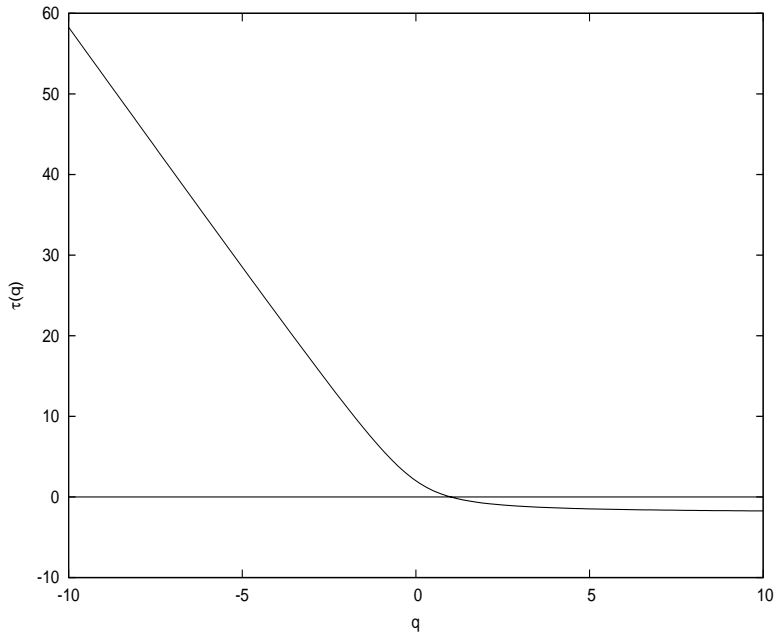}
\end{center}
\caption{The plots of $\tau(q)$ vs $q$ to show its slope as a function of $q$ varies.
}
\label{fig:fg5}
\end{figure}

Of all the conservation laws we find that $M_3(t)=\sum_ix_i^3$ is a special one since we can use it as
a multifractal measure consisting of members $p_i$, the fraction of the total measure $p_i=x_i^3/\sum_ix_i^3$, distributed on
the geometric support WPSL.
That is, we assume that the $i$th block is occupied with cubic power of its own length $x_i$.  
The corresponding "partition function" of multifractal formalism then is  \cite{ref.stanley}
\begin{equation}
\label{eq:10}
Z_q(t)=\sum_ip_i^q\sim M_{3q}(t).
\end{equation}
Its solution can immediately be obtained from equation (\ref{eq:qmoment}) to give
\begin{equation}
\label{eq:11}
Z_q(t)\sim t^{\{\sqrt{9q^2+16}-(3q+2)\}/2}.
\end{equation} 
Using the square root of the mean area
$\delta(t)=\sqrt{M(2,2;t)/M(1,1;t)}\sim t^{-1/2}$ as the yard-stick to express the partition function $Z_q$
gives the weighted number of squares $N(q,\delta)$ needed to cover the measure which we find 
decays following power-law
\begin{equation}
\label{weightednumber}
N(q,\delta)\sim \delta^{-\tau(q)},
\end{equation}
where the mass exponent 
\begin{equation}
\label{massexponent}
\tau(q)=\sqrt{9q^2+16}-(3q+2).
\end{equation} 
The non-linear nature of $\tau(q)$, see figure 5 for instance, suggests
that the gap exponent
\begin{equation}
\label{eq:12}
\Delta=\tau(q)-\tau(q-1)
\end{equation}
is different for every $q$ value. It implies that we require an infinite hierarchy of
exponents to specify how the moments of the probabilities $\{p\}$s 
scales with $\delta$. Now if we choose $q=0$ then it gives an estimate of
the number of squares $N(0,\delta=N(\delta)$ of sides $\delta$ we need to cover the support on which the 
members of the population is distributed. We find that $N(\delta)$ scales as 
\begin{equation}
\label{eq:13}
N(q=0,\delta)\sim \delta^{-\tau(0)},
\end{equation}
where $\tau(0)=2$ is the Hausdorff-Besicovitch dimension of the support 
\cite{ref.multifractal_1}. On the other hand,
if we choose $q=1$ we have $Z_1=\sum_i p_i=const.$ and hence we must have $\tau(1)=0$. This is indeed
the case according to equation (\ref{massexponent}). Therefore, $\tau(0)=2$ (the dimension of the support) and
$\tau(1)=0$ (required by the normalization condition) 
are often considered as the first self-consistency check for the multifractal analysis.

The role of $q$ in the partition function $Z_q$ can be best understood by appreciating the fact that
the large values of $q$ favours contributions in the $Z_q$ from the blocks with relatively high values of $p_i$  
since $p_i^q>>p_j^q$ with $p_i>p_j$ provided $q>>1$. On the other hand, $q<<-1$ favours contributions
in the $Z_q$ from those blocks which are occupied with relatively low values of the measure $p_i$.
For further elucidation we find it worthwhile to consider the slope of the curve $\tau(q)$ vs $q$ which is given by
\begin{equation}
\label{eq:14}
{{d\tau(q)}\over{dq}}=-\lim_{\delta \rightarrow 0}{{\sum_ip_i^q\ln p_i}\over{(\sum_ip_i^q)\ln \delta}}.
\end{equation}
Now $\tau(q)$ vs $q$ curve has the maximum slope in the limit $q\rightarrow -\infty$ and hence we can write
\begin{equation}
\label{eq:15}
{{d\tau(q)}\over{dq}}=-\alpha_{{\rm max}}.
\end{equation}
It implies that the right hand side of equation (\ref{eq:14}) is dominated
by $p_{{\rm min}}$, the minimum of $p_i$, in the sum and hence we have
\begin{equation}
\label{eq:16}
{{d\tau(q)}\over{dq}}\Big |_{q\rightarrow -\infty}=- \lim_{\delta \rightarrow 0}{{\ln p_{{\rm min}}}\over{\ln \delta}}
=-\alpha_{{\rm max}},
\end{equation}
and hence we get
\begin{equation}
p_{{\rm min}}\sim \delta^{\alpha_{{\rm max}}}.
\end{equation}
A similar argument in the limit $q \rightarrow \infty$ leads to the conclusion that the minimum slope of the 
$\tau(q)$ vs $q$ curve is given by 
\begin{equation}
\label{eq:17}
 {{d\tau(q)}\over{dq}}\Big |_{q\rightarrow +\infty}=-\lim_{\delta \rightarrow 0}{{\ln p_{{\rm max}}}\over{\ln \delta}}
=-\alpha_{{\rm min}},
\end{equation}
 where $p_{{\rm max}}$ is the largest value of $p$  which gives the minimum value of $\alpha$ and hence
\begin{equation}
p_{{\rm max}}\sim \delta^{\alpha_{{\rm min}}}.
\end{equation}
So, in general
we can write
\begin{equation}
\label{eq:18}
{{d\tau(q)}\over{dq}}=-\alpha(q),
\end{equation} 
where the exponent $\alpha(q)$ is known as the Lipschitz-H\"{o}lder exponent and
\begin{equation}
\label{eq:19}
p\sim \delta^\alpha.
\end{equation}


\begin{figure}[ht]
\begin{center}
\includegraphics[width=12.5cm,height=8.5cm]{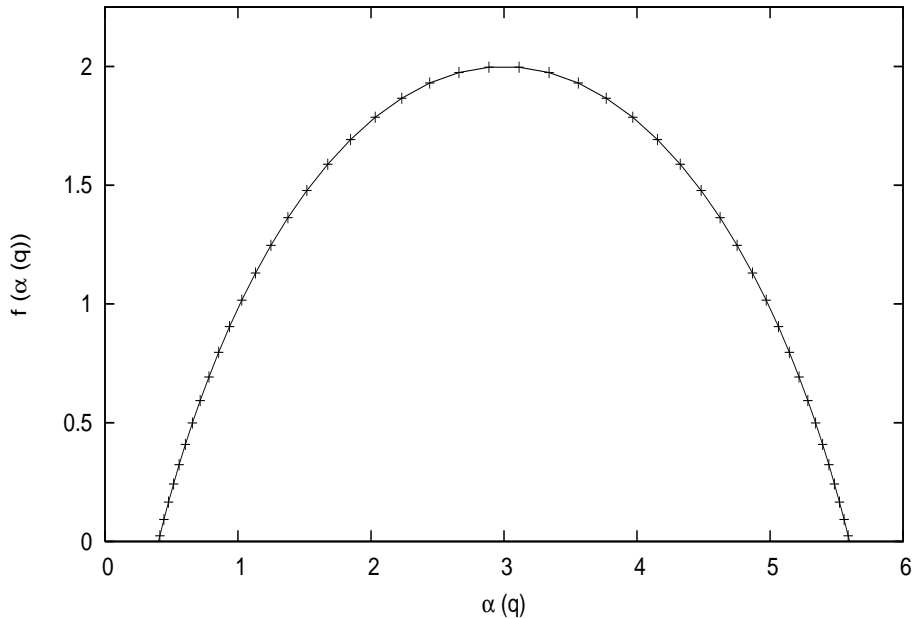}
\end{center}
\caption{The $f(\alpha)$ spectrum.
}
\label{fig:fg6}
\end{figure}

We now perform the Legendre transformation of the mass exponent $\tau(q)$ by using the Lipschitz-H\"{o}lder exponent 
given equation (\ref{eq:18})
as an independent variable to obtain the new function 
\begin{equation}
\label{eq:21}
f(\alpha)=q\alpha+\tau(q).
\end{equation}
Replacing $\tau(q)$ in equation (\ref{weightednumber}) in favour of $f(\alpha)$ we find that 
\begin{equation}
\label{eq:22}
N(q(\alpha),\delta) \sim \lim_{\delta\rightarrow 0}\delta^{q\alpha-f(\alpha)}.
\end{equation}
On the other hand, using $p\sim \delta^\alpha$ in the expression for partition
function $Z_q=\sum_ip^q$ and replacing the sum by integral while indexing the blocks by continuous 
Lipschitz-H\"{o}lder exponent $\alpha$ as variable with a weight $\rho(\alpha)$ we obtain   
\begin{equation}
\label{eq:23}
N(q(\alpha),\delta) \sim \int \rho(\alpha)d\alpha N(\alpha,\delta)\delta^{q\alpha},
\end{equation}
where $N(\alpha,\delta)$ is the number of squares of side $\delta$ needed to cover the measure indexed by 
$\alpha$. Comparing Eqs. (\ref{eq:22}) and (\ref{eq:23}) we find 
\begin{equation}
\label{eq:24}
N(\alpha,\delta)\sim \delta^{-f(\alpha)}.
\end{equation}
It implies that we have a spectrum of spatially intertwined fractal 
dimensions 
\begin{equation}
f(\alpha(q))={{16}\over{\sqrt{9q^2+16}}}-2,
\end{equation}
are needed to characterize the measure. 
That is, the size disorder of the blocks are multifractal in character since the measure $\{p_\alpha\}$ is related
to size of the blocks.  
That is, the distribution of $\{p_\alpha\}$ in WPSL can
be subdivided into a union of fractal subsets each with fractal dimension $f(\alpha)\leq 2$ in which the 
measure $p_\alpha$ scales as $\delta^\alpha$.
Note that $f(\alpha)$ is always concave in character (see figure 6) with a single maximum at $q=0$ which 
corresponds to the dimension of the WPSL with empty blocks.

On the other hand, we find that the entropy $S(\delta)=-\sum_i p_i\ln p_i$ associated with the partition of the 
measure on the support (WPSL) by 
using the relation $\sum_i p_i^q\sim \delta^{-\tau(q)}$ in the definition of $S(\delta)$. Then  
a few steps of algebraic manipulation reveals that $S(\delta)$ 
exhibits scaling 
\begin{equation}
S(\delta)=\ln\delta^{-\alpha(1)}
\end{equation} 
where the exponent $\alpha_1={{6}\over{5}}$ obtained from 
\begin{equation}
\alpha(q)= -\left. d\tau(q)/dq \right |_{q}.
\end{equation} 
It is interesting to note that $\alpha(1)$ is related to the generalized dimension $D_q$, also related to the R\'{e}nyi entropy 
$H_q(p)={{1}\over{q-1}}\ln \sum_i p_i^q$ in the 
information theory, given by
\begin{equation}
\label{dq}
D_q=\lim_{\delta\rightarrow 0} \Big [{{1}\over{q-1}}{{\ln \sum_i p_i^q}\over{\ln\delta}}\Big ]={{\tau(q)}\over{1-q}},
\end{equation}
which is often used in the multifractal formalism as it can also provide insightful interpretation. 
For instance, $D_0=\tau(0)$ is the dimension of the support, $D_1=\alpha_1$ is the Renyi information dimension 
and $D_2$ is known as the correlation dimension \cite{ref.procaccia,ref.renyi_entropy}. 
Multifractal analysis was initially proposed to treat
turbulence but later successfully applied in a wide range of exciting field of research.
For instance, it has been found recently that the wild fluctuations of the wave functions at the Anderson and the quantum Hall 
transition can be best described by multifractality \cite{ref.mandelbrot,ref.anderson}. 
Recently, though it has got renewed momentum
as it has been found that the probability density function
at the Anderson and the quantum Hall transition exhibits multifractality since in the vicinity of the transition point
fluctuations are wild - a characteristic feature of multifractal behaviour \cite{ref.anderson}.

In an attempt to understand the origin of multifractality we now consider the case 
where the generator divides the initiator into four equal blocks instead of randomly into four blocks.
If the generator is applied over and over again thereafter to only one of the available squares by picking preferentially 
with respect to their areas then it results in the kinetic square lattice (KSL). 
Within the rate equation approach it can be described by the kernel 
\begin{equation}
F(x_1,x_2,y_1,y_2)=(x_1+x_2)(y_1+y_2)\delta(x_1-x_2)\delta(y_1-y_2).
\end{equation}
and hence the resulting rate equation can be obtained after substituting it in equation (\ref{eq:1}) to give
\begin{equation}
\label{eq1}
{{\partial C(x,y;t)}\over{\partial t}} = -{{1}\over{4}}xyC(x,y;t)+ 
4^2xyC(2x,2y;t).
\end{equation}
Incorporating equation (\ref{eq4}) in equation (\ref{eq1}) yield
\begin{equation}
\label{momenteq_2}
{{dM(m,n;t)}\over{dt}}=-\Big ( {{1}\over{4}}-{{4}\over{2^{m+n}}}\Big )M(m+1,n+1;t).
\end{equation}
To obtain the solution of this equation in the long-time limit, 
we assume the following power-law asymptotic behaviour of $M(m,n;t)$ and write
\begin{equation}
M(m,n;t)\sim A(m,n)t^{\theta(m+n)},  
\end{equation}
with $\theta(4)=0$ since total area obtained by setting $m=n=2$ is an obvious conserved quantity. 
Using it in equation (\ref{momenteq_2}) yields the following difference equation
\begin{equation}
\theta(m+n+2)=\theta(m+n)-1.
\end{equation}
Iterating it subject to the condition that $\theta(4)=0$ gives
\begin{equation}
M(m,n;t)\sim t^{-{{m+n-4}\over{2}}}.
\end{equation}
Apparently it appears that in addition to the conservation of the total area $M(2,2;t)$, we find that the
integrals $M(3,1;t)$ and $M(1,3;t)$) are also conserved. Interestingly, all the three integrals 
$M(2,2;t)$, $M(3,1;t)$ and $M(1,3;t)$ effectively describe the same physical quantity since all the blocks 
are square in shape and hence 
\begin{equation}
\sum_{i=1}^N x_i^2=\sum_{i=1}^N y_i^2=\sum_{i=1}^N x_iy_i.
\end{equation}
Therefore, in reality the system obeys only one conservation law - conservation of total area.

We again look into the $q$th moment of $n(x,t)$ using equation (\ref{eq:8}) and appreciating the fact
that $M_q(t)$ equal to $M(q+1,1;t)$ or $M(1,q+1;t)$ we immediately find that 
\begin{equation}
\label{eq:ksl_moment_2}
M_q(t)\sim t^{-{{q-2}\over{2}}}.
\end{equation} 
Unlike in the previous case where the exponent of the power-law solution of the $M_q(t)$ is non-linear, here 
we have an exponent which is linear in $q$. It immediately implies that in the case kinetic square lattice
\begin{equation}
<x^q>=<x>^q,
\end{equation}
and hence a single length-scale is enough to characterize all the moments of $n(x,t)$. That is, the system now
exhibits simple scaling instead of multiscaling. Like before let us consider that each block is occupied with 
a fraction of the measure equal to square of its own length or area $p_i=\sum_i^Nx_i^2$ and hence
the corresponding partition function is
\begin{equation}
Z_q=\sum_i^N p_i^q=M_{2q}(t).
\end{equation}
Using equation (\ref{eq:ksl_moment_2}) we can immediately write its solution
\begin{equation}
Z_q(t)\sim t^{-{{2q-2}\over{2}}}.
\end{equation}
Expressing it in terms of the square root of the average area $\delta\sim t^{-1/2}$ gives the 
weighted number of square $N(q,\delta)$ of side $\delta$ needed to cover the measure which has the following 
power-law solution
\begin{equation}
N(q,\delta)\sim \delta^{-(2-2q)},
\end{equation}
where mass exponent $\tau(q)$ is
\begin{equation}
\tau(q)=2-2q.
\end{equation}
The Legendre transform of the mass exponent is a constant
\begin{equation}
f(\alpha)=2,
\end{equation}
and so is the generalized dimension  
\begin{equation}
D_q=2.
\end{equation}
We thus find that if the generator divides the initiator into four equal square and we apply it thereafter
sequentially then the resulting lattice is no longer a multifractal. The reason is that the distribution of the 
population in the resulting support in this case is uniform. The two model therefore provides a unique opportunity 
to look for possible origin of multifractality. The two models discussed in this talk differs only in the definition
of the generator. In the case when the generator divides the initiator randomly into four blocks and we apply it
over and over again sequentially then we have multifractality since the underlying mechanism in this case is governed by random
multiplicative process. This is not, however, the case if the generator divides the initiator into
equal four blocks and apply it over and over again sequentially since the resulting dynamics is governed by
deterministic multiplicative process instead.

\begin{figure}[ht]
\begin{center}
\includegraphics[width=12.5cm,height=8.5cm]{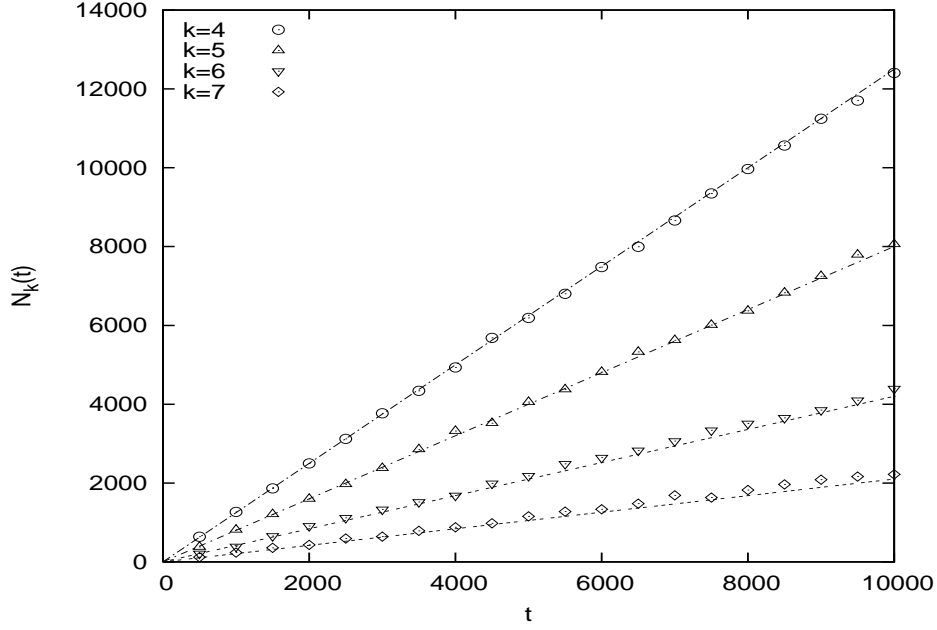}
\end{center}
\caption{Shown is the growth of the number of blocks $N_k(t)$ with exactly $k=4,5,6,7$
neighbours as a function of time.
}
\label{fig:fg7}
\end{figure}

\begin{figure}[ht]
\begin{center}
\includegraphics[width=12.5cm,height=8.5cm]{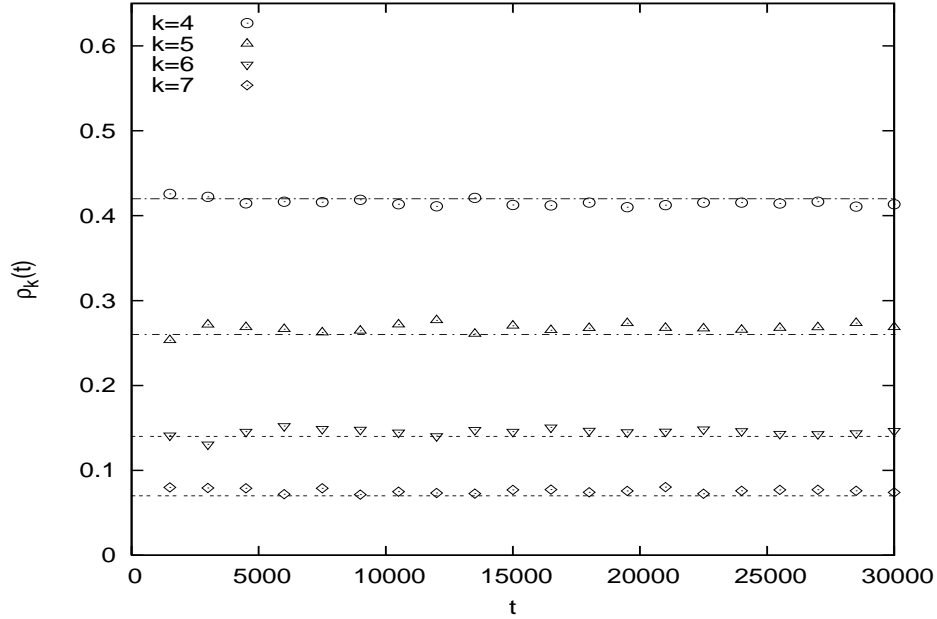}
\end{center}
\caption{Fraction of the blocks $\rho_k(t)$ with $k=4,5,6,7$ neighbours are drawn as a function of time.
}
\label{fig:fg8}
\end{figure}

\section{Scale-free properties of WPSL}

Defining each step of the algorithm as one time unit and imposing periodic boundary condition in the simulation
we can immediately write an exact expression for the total number of blocks as a function of time: $N(t)=1+3t$. However, unlike the square lattice 
where every block shares its borders exactly with four neighbours, the blocks in the WPSL can
have their common borders with at least four or more blocks albeit all the blocks have exactly four sides. 
In fact, we can characterize individual 
blocks of the WPSL by their coordination numbers, which is defined as the number of neighbours with whom 
they have common borders, which
we find neither a constant like the square lattice nor has a typical mean value like the Voronoi lattice. Instead,
the coordination number that each block assumes in the WPSL is random and evolves with time.  
The coordination number disorder in the WPSL can, therefore, be regarded as of annealed type. 
Our data extracted from numerical simulation suggest  
that the number of blocks $N_k(t)$ which have coordination number $k$ (or $k$ neighbours) continue to
grow linearly with time $N_k(t)=m_k t$ (see figure 7). On the other hand, the number of total blocks $N(t)=\sum_k N_k(t)$ 
at time $t$ in the lattice also grow linearly with time $N(t)=1+3t$ and hence $N(t)\sim 3t$ 
in the long-time limit. The ratio of the two quantities $\rho_k(t)=N_k(t)/N(t)$ that
describes the fraction of the total blocks which have coordination number $k$ is
$\rho_k(t)=m_k/3$ and it is clearly independent of time {\it vis-a-vis} of the size of the lattice
(see figure 8). 
It implies that we can take a lattice of 
sufficiently large size and study its properties as these properties without being worried about its exact size.

\begin{figure}[ht]
\begin{center}
\includegraphics[width=12.5cm,height=8.5cm]{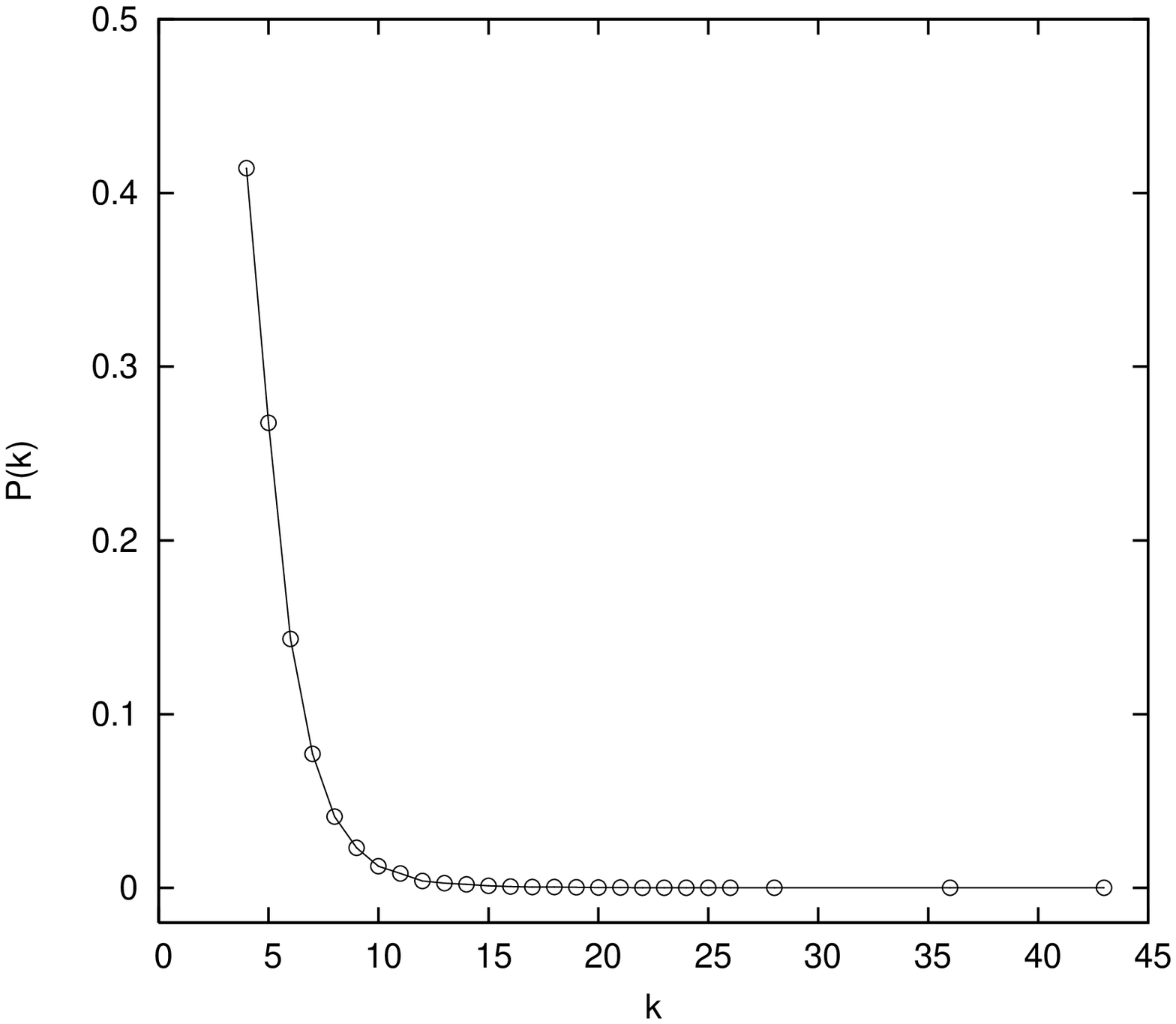}
\end{center}
\caption{Shown is the coordination number distribution function $P(k)$ vs $k$ which is also
equivalent to the
degree distribution of the DWPSL.
}
\label{fig:fg9}
\end{figure}

\begin{figure}[ht]
\begin{center}
\includegraphics[width=12.5cm,height=8.5cm]{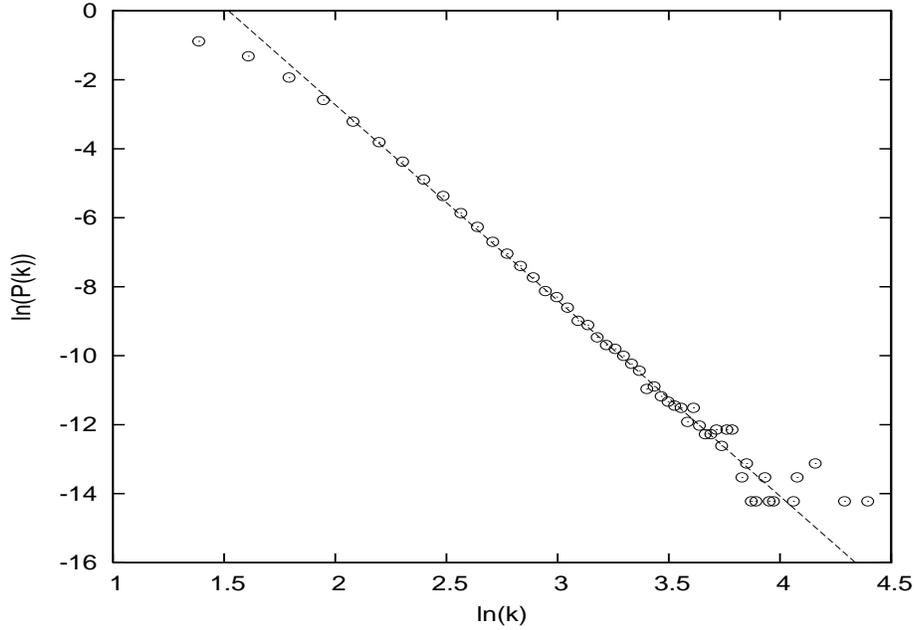}
\end{center}
\caption{Drawn is plot of $\ln (P(k))$ against $\ln (k)$
for the DWPSL network where data points
represent average of $50$ independent realizations.
The line have slope exactly equal to $5.66$ revealing power-law degree distribution with exponent $\gamma=5.66$.
}
\label{fig:fg10}
\end{figure}

We now assume that that blocks in the WPSL are equivalent regardless of their sizes and 
then ask: 
What is the probability that a block picked at random with equal {\it a priori} 
probability from all the available blocks of the lattice have coordination number $k$?
Answer to this question lies in collecting data for $\rho_t$ as a function of $k$ and plotting a normalized histogram
which effectively gives the coordination number distribution function $\rho_t(k)\equiv P(k)$ where subscript
$t$ indicates fixed time.  
We first plot $P(k)$ in figure 9 to have a glimpse of its behaviour as a function of $k$. A closer look into the data reveals
that there exist scarce data points near the tail. The long tail with scarce data points turns into a noisy or 
fat-tail when we plot $\ln P(k)$ vs $\ln (k)$. This is drawn in figure 10 
where data points represents
ensemble average over $50$ independent 
realizations and it clearly suggests that fitting a straight line near the tail is possible. It implies that the 
coordination number distribution decays obeying power-law 
\begin{equation}
\label{degreedistribution}
P(k)\sim k^{-\gamma}, 
\end{equation}
with heavy or fat-tail reflecting scarce data points near the tail-end of figure 10.
However, the noise in the tail-end complicate the process of identifying the range over which the
power-law holds and hence of estimating the exponent $\gamma$. One way of reducing the 
noise at the tail-end is to plot   
cumulative distribution $P(k^\prime \geq k)$ which is related to degree distribution $P(k)$ via
\begin{equation}
P(k)=-{{dP(k^\prime\geq k)}\over{dk}}.
\end{equation}
We therefore plot  $\ln (P(k^\prime >k))$ vs $\ln (k)$ in figure 11 
using the same data of figure 10 
and find that the heavy tail smoothes out naturally where no data is obscured. 
The straight line fit of figure 11 has a slope $\gamma-1=4.66$ which 
indicates that the degree distribution (figure 10) decays following power-law with exponent $\gamma=5.66$.
A power-law distribution function is often regarded as scale-free since it looks the same 
whatever scale we look at it. 

\begin{figure}[ht]
\begin{center}
\includegraphics[width=12.5cm,height=8.5cm]{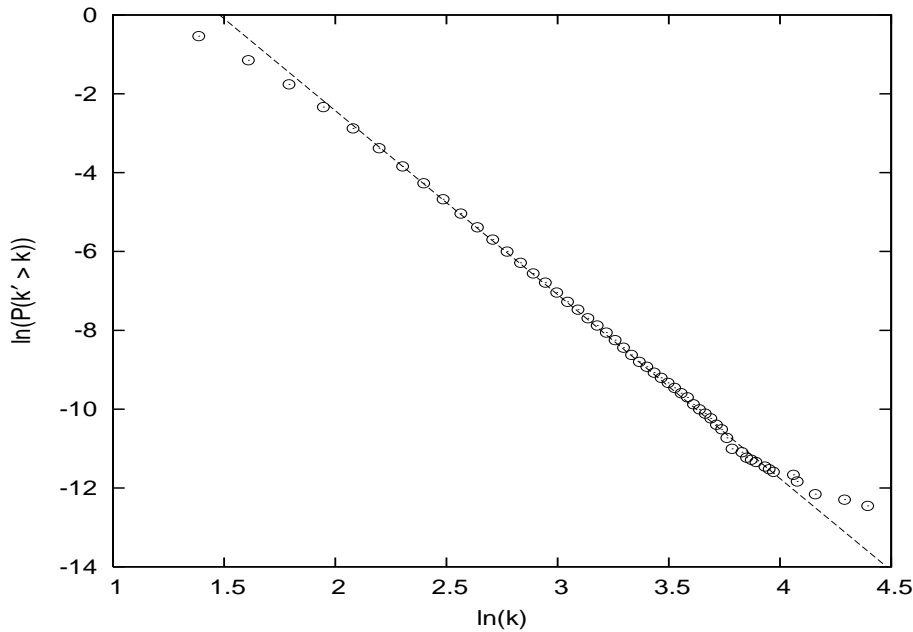}
\end{center}
\caption{Cumulative degree distribution $P(k^\prime>k)$ is shown using the same data as of figure 4.
The dotted line with slope equal to $\gamma-1=4.66$ is drawn to guide our eyes.
}
\label{fig:fg11}
\end{figure}

We thus find that in the large-size limit 
the WPSL develops some order in the sense that its annealed coordination number disorder is scale-free 
in character. This is in sharp contrast to the quenched coordination number disorder found in the Voronoi lattice where 
it is almost impossible to find cells which have
significantly higher or fewer neighbours than the mean value $k=6$ \cite{ref.vd}.
In fact, it has been shown that 
the coordination number distribution of the Voronoi lattice is Gaussian in character instead. 
The power-law coordination number distribution in the WPSL is reminiscent of the scale-free degree distribution 
of complex network. The past decade have witnessed a surge of interest in the theory of complex network 
resulting in the dramatic advances in the field thanks to the seminal work of A.-L. Barabasi and his co-workers.
It is worth mentioning that the dual of the WPSL, obtained replacing each block with a node at its centre and
common border between blocks with an edge joining the two nodes (see figure 12 where we have illustrated the
network aspect of the WPSL) is a network whose degree distribution has the same data as the coordination number
distribution of the WPSL. 

\begin{figure}[ht]
\begin{center}
\includegraphics[width=12.5cm,height=8.5cm]{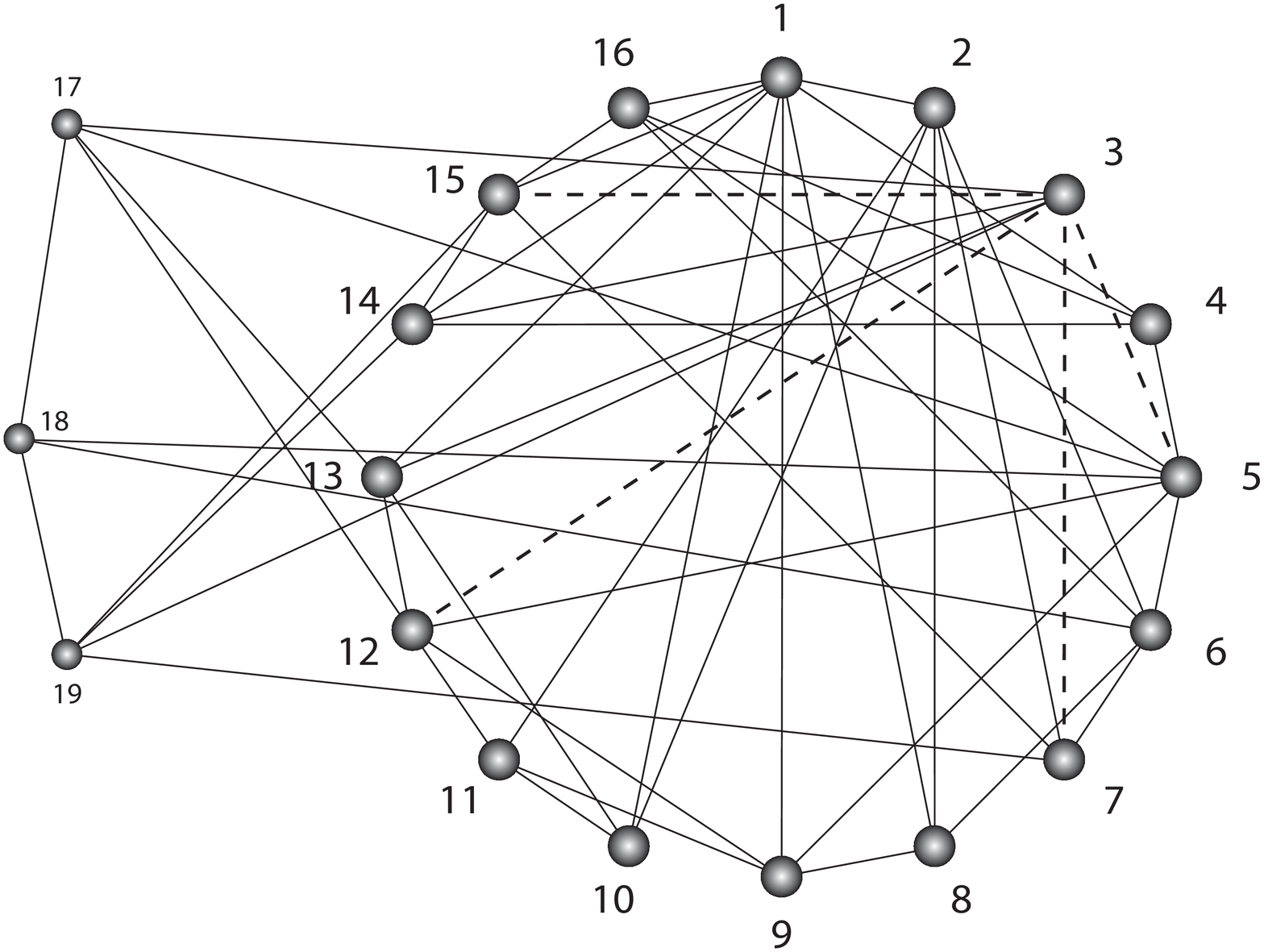}
\end{center}
\caption{Nodes in the circle and their links with other nodes illustrate the network topology at step $5$ of Fig. 2.
New nodes ($17$, $18$ and $19$) and links
to be added and links to be removed (broken lines) in step $6$ again of Fig. 2 are also shown to illustrate
the dynamics.
}
\label{fig:fg12}
\end{figure}
  
The square lattice, on the other hand, is self-dual and its degree distribution is $P(k)=\delta(k-4)$.
Further, it is interesting to point out that the exponent $\gamma=5.66$ is significantly higher than 
usually found in most real-life network which is typically $2<\gamma\leq 3$. This suggests that in addition to the PA rule the network
in question has to obey some constraints. For instance, nodes in the WPSL are spatially embedded in Euclidean space,
links gained by the incoming nodes are constrained by the spatial location and the fitting parameter of the nodes.  
Owing to its unique dynamics this was not unexpected.
Perhaps, it is noteworthy to mention that the degree distribution  of the electric power grid, whose nodes like WPSL are also embedded in
the spatial position, is shown to exhibit power-law but with exponent $\gamma_{{\rm power}}=4$  \cite{ref.barabasi}.

The power-law degree distribution $P(k)$ has been found
in many seemingly unrelated real life networks. It implies that there must exists some common 
underlying mechanisms for which disparate systems behave in such a 
remarkably similar fashion \cite{ref.review}. Barabasi and Albert argued that the growth and the PA rule are the 
main essence behind the emergence of such power-law. Indeed, the DWPSL network too grows with time but in sharp contrast to the BA model 
where network grows by addition of one single node with $m$ edges per unit time,
the DWPSL network grows by addition of a group of three nodes which are already linked by two edges. 
It also differs in the way incoming nodes establish links with the existing nodes.

To understand the growth mechanism of the DWPSL network let us look into the $j$th step of the algorithm. First a node, say it is labeled as $a_p$, is picked from the $(3j-2)$ 
nodes preferentially with respect to the fitness parameter of a node (i.e., according to their respective areas). Secondly, connect the node $a_p$
with two new nodes $(3j-1)$ and $(3j+1)$ in order to establish their links with the existing network.
At the same time, at least two or more links of $a_p$ with other nodes are removed (though the exact number depends on the number of neighbours $a_p$ already 
has) in favour of linking them among the three incoming nodes in a self-organized fashion. 
In the process, the degree $k_p$ of the node $a_p$ will either decrease (may turn into a node with marginally low 
degree in the case it is highly connected node) or at best remain the same but will never increase. 
It, therefore, may appear that the PA rule is not followed here. 
A closer look into the dynamics, however, reveals otherwise. It is interesting to note that an existing nodes during the process gain links
only if one of its neighbour is picked not itself. It implies that the higher the links (or degree) a node has, the higher 
its chance of gaining more links since they can be reached in a larger number of ways. It essentially embodies the intuitive idea of PA rule.
Therefore, the DWPSL network can be seen to follow preferential attachment rule but in disguise.

\section{Summary}

To summarize, we proposed a model that generates weighted planar stochastic lattice (WPSL) which emerges through evolution.
Unlike the regular lattice where every block or cell
has exactly the same coordination number, the WPSL is seemingly disordered both in terms of the
coordination number and in terms of its block size distribution. One of our primary goal
was to find some order in such seemingly 
disordered system since finding order is always an attractive proposition for physicists.
To this end, our first attempt was to quantify the size disorder of the lattice and we found several interesting results.
One of the most interesting results that we found is that $\sum_i^N x_i^{n-1} y_i^{4/n-1}$ remains a constant 
$\forall \ n$ regardless of the size of the lattice where
blocks are labeled as $i=1,2,...,N$ and $x_i$ and $y_i$ are their  respective 
length and width. Yet another interesting fact is that the numerical
values of the conserved quantities $\sum_i^N x_i^{n-1} y_i^{4/n-1}$ for a given value of $n$ except $n=2$ 
are never the same in different realizations which is clearly an indication of wild fluctuation. 
Of the conservation laws, we found 
a special one obtained by choosing either $n=1$ or $n=3$ to give $\sum_{i=1}^N y_i^3$ or  $\sum_{i=1}^N x_i^3$ which
can be picked as a measure. It implies that if the blocks are populated with a fraction of the measure equal to cubic power 
of their respective length or width then its distribution on the WPSL is multifractal in nature 
- a property that quantifies the wild fluctuation of block size distribution of the WPSL. That is, the probability
distribution of $x_i^3$ is multifractal in the sense that a single exponent is not enough to characterize its distribution
on the support, namely the WPSL, instead we need a spectrum of exponent $f(\alpha)$. We derived an exact 
expression for the $f(\alpha)$ spectrum and obtained information dimension $D_1=6/5$ with which the entropy of the 
measure scales $S(\delta)\sim \ln \delta^{-D_1}$. 
To look for a
possible origin of multifractality we also studied the kinetic square lattice which is actually the deterministic counterpart
of the WPSL. It led to the conclusion that as soon as randomness is ceased from the definition of the generator
the resulting lattice can no longer be quantified as multifractal rather it possesses all the properties of the 
square lattice but only in the long-time limit. 

Our second attempt
was to quantify the coordination number disorder.  We have shown numerically that the coordination number disorder is scale-free in character 
since the degree distribution of its dual (DWPSL), which is topologically identical to the network obtained
by considering blocks of the WPSL as nodes and the common border between blocks as links, exhibits power-law.  
In other words, the coordination number distribution function of the WPSL decays following power-law
with exponent $\gamma=5.66$. However, the novelty of this network is that it grows by addition of a group of 
already linked nodes which then establish links with the existing nodes
following PA rule, though in disguise, in the sense that existing nodes gain links only if one of their neighbour 
is picked, not the node itself. 
Finally, such multifractal lattice with scale-free coordination disorder can be of great interest as it has the potential to mimic 
disordered medium on which one can study various physical phenomena like percolation and random walk problems etc. 
One may also study the phase transition and critical behaviour in such multifractal scale-free lattice. Indeed, phase
transition and critical behaviour in complex network have attracted much of our recent attention. The advantage of the
WPSL over many 
known complex networks is that its nodes or blocks are embedded in spatial position. That is, one may assume that
interacting particles located on the sites with great many different neighbours and therefore its result must 
differ from that of a system where sites are located on a regular lattice. We intend to work in these directions in our 
future endeavour.

NIP gratefully acknowledges support from the Bose Centre for Advanced Study and Research in Natural Sciences.

\end{document}